\documentclass[11pt, a4paper]{article}

\usepackage[margin=1in]{geometry}
\usepackage{xcolor}
\usepackage{hyperref}
\usepackage{setspace}

\usepackage{graphicx}
\usepackage{enumitem}
\usepackage{float}

\begin{document}

{\LARGE \textbf{An agentic framework for gravitational-wave counterpart association in the multi-messenger era}}\\[1em]

\textbf{Yiming Dong,\textsuperscript{1,2} Yacheng Kang,\textsuperscript{1,2} Junjie Zhao,\textsuperscript{3} Xinyuan Zhu,\textsuperscript{4} Ziming Wang,\textsuperscript{1,2} and Lijing Shao\textsuperscript{2,5*}}

\textsuperscript{1}Department of Astronomy, School of Physics, Peking University, Beijing 100871, China\\
\textsuperscript{2}Kavli Institute for Astronomy and Astrophysics, Peking University, Beijing 100871, China\\
\textsuperscript{3}Institute for Gravitational Wave Astronomy, Henan
Academy of Sciences, Zhengzhou 450046, China\\
\textsuperscript{4}School of Information Science and Technology, University of Science and Technology of China, Hefei 230026, China\\
\textsuperscript{5}National Astronomical Observatories, Chinese Academy
of Sciences, Beijing 100101, China\\
\textsuperscript{*}Correspondence: lshao@pku.edu.cn

\textbf{SUBJECT AREA:} Astrophysics\\

As a landmark discovery in astronomy and fundamental physics, GW170817 was the
first binary neutron star (BNS) merger detected through gravitational waves
(GWs) and accompanied by multi-wavelength electromagnetic (EM)
counterparts.\textsuperscript{1}  Joint observations of the GW
signal, the associated short-duration gamma-ray
burst (SGRB) GRB, and the kilonova
transient, have yielded remarkable scientific
returns,\textsuperscript{1} including a smoking-gun evidence for the BNS merger origin of SGRBs and
kilonovae,\textsuperscript{2} and an independent measurement of the Hubble
constant via the standard siren method.\textsuperscript{3}
Besides BNSs, neutron star--black hole (NS--BH) mergers, as well as stellar-mass binary black hole (BBH) mergers in active galactic nucleus (AGN) disks, have also been speculated to give rise to EM counterparts.\textsuperscript{4} Unfortunately, GW follow-up
campaigns have so far been unsuccessful to   identify convincing EM counterparts
to NS--BH or BBH mergers. The non-detections may result from the large distances
of these events, incomplete coverage of their large GW localization regions, or
intrinsically fainter EM emissions from them.

With the advent of next-generation GW detectors and advancing EM facilities, multi-messenger astronomy is expected to boost with richer
scientific outcomes for all types of compact binary mergers discussed above.
These opportunities, however, are accompanied by greater data-analysis
challenges. In particular, searching for association between GW signals and
their EM counterparts is vital for enabling subsequent multi-messenger studies.
The expected increase in co-detecting them will place growing pressure on both
computational resources and human effort, preventing us to fully exploit the
scientific value of data.

The rapid advancement of artificial intelligence provides a promising pathway to
help address emerging challenges. Machine learning has enabled real-time
parameter estimation for BNS merger signals, delivering accurate estimates of
sky localization within seconds,\textsuperscript{5} thereby supporting early
alerts for multi-messenger follow-up. It has also been applied to searches for
EM counterparts, assisting candidate prioritization in complex observational
environments.\textsuperscript{6} Beyond conventional machine-learning
approaches, the recent rise of large language models (LLMs) further show the
broad potential of artificial intelligence for scientific discoveries. Owing to
their rapidly-advancing reasoning capabilities, LLMs have been successfully
explored in diverse scientific tasks, including 
automated scientific discovery.\textsuperscript{7} These
LLM systems are able to achieve robust performance while substantially reducing
the human effort and alleviating the pressure imposed by the rapidly growing
data volume.

Motivated by this, we develop \textbf{GW-Eyes}, a first LLM-powered agentic
framework designed to automate GW counterpart association tasks while assisting
human experts through natural-language interaction with auxiliary functions such
as catalog management, skymap visualization, and rapid verification. GW-Eyes
integrates domain-specific tools tailored to the distinctive data structures of
multi-messenger observations and dynamically retrieves and updates the latest
information from the internet. We construct simulated datasets to evaluate the
accuracy and efficiency of its tool use, and further demonstrate its application
to EM counterpart searches using real observational catalogs. Through GW-Eyes,
we demonstrate how LLMs can participate in future multi-messenger observations
and help address challenges in the big-data era.


\section*{FRAMEWORK OVERVIEW}
\label{ sec:overview }

\begin{figure}[H]
    \centering
    \includegraphics[width=1\textwidth]{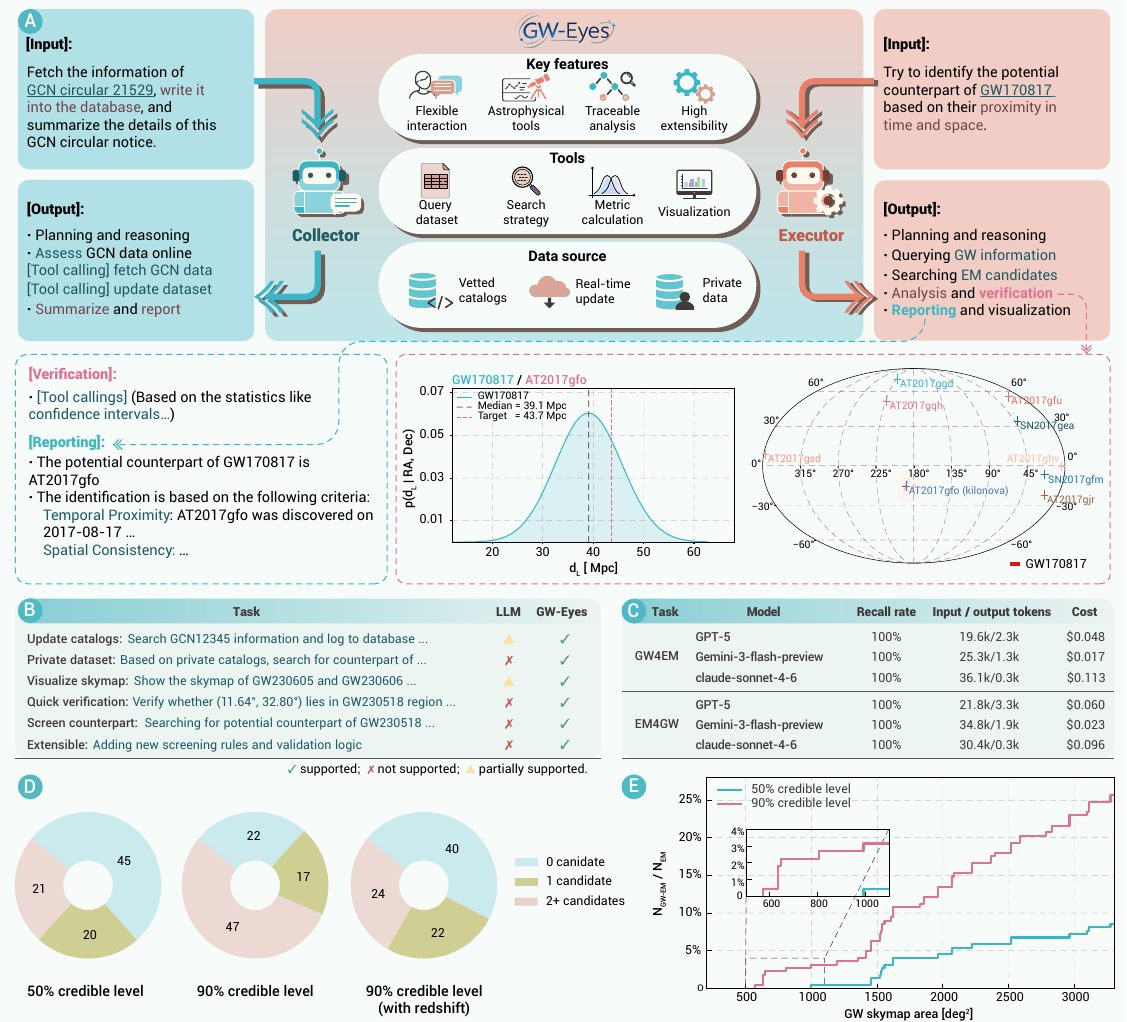}
    \caption{\textbf{GW-Eyes enables autonomous GW counterpart association in
    multi-messenger astronomy.} (A) Based on a Collector-Executor sub-agent
    architecture with domain-specific tool integration, GW-Eyes performs
    counterpart association and related tasks through natural-language
    interaction. (B) Compared with general-purpose LLMs, GW-Eyes achieves
    improved adaptation to the specialized domain of multi-messenger astronomy.
    (C) GW-Eyes demonstrates stable performance and lower cost in evaluations on
    simulated datasets. (D) Using real catalogs, AGNFRC and GWTC-4.0, GW-Eyes
    performs preliminary associations based on temporal and spatial
    correlations. (E) The cumulative ratio $N_{\rm GW-EM} / N_{\rm EM}$ as a
    function of the GW skymap area threshold provides statistical insight into
    counterpart association and shows the growing data pressure expected from
    the increasing rate of future closely detected events.}
\label{fig: total}
\end{figure}

GW-Eyes is a general LLM-driven agentic system designed for GW counterpart association.
As illustrated in Figure~\ref{fig: total}A, it comprises two sub-agents,
Collector and Executor, and integrates domain-specific tools through the Model
Context Protocol.  Collector is responsible for data acquisition and initial
preprocessing, including the retrieval of GW posterior information, public EM
event catalogs, and real-time circulars from General Coordinates Network (GCN).
Executor operates on the database to perform counterpart association tasks, as
well as auxiliary functions such as visualization and rapid verification.

Upon receiving a data-acquisition task, such as updating GCN circular records,
Collector first retrieves the relevant online GCN circular contents (see
Figure~\ref{fig: total}A). These records are typically human-generated text,
where key event information, such as sky position and redshift, is embedded in
natural language.  Collector then automatically interprets and extracts the
information relevant to counterpart association, performs rule-based and optional self-consistency
checks, and stores it in the database
according to predefined rules and thereby enabling subsequent analyses by the
Executor. 

For counterpart association tasks, such as searching for EM counterpart
candidates to a given GW event, Executor first decomposes the request and
formulates an analysis strategy.  It then invokes tools to access GW posteriors
and EM catalogs, and evaluates their spatial associations using statistical
metrics. After generating a list of candidates, Executor can provide
visualizations, including sky-localization maps and the luminosity distance
posterior, together with a detailed report summarizing the selection criteria
and final results.  The association analysis can be performed in either
direction: starting from a GW event to identify potential EM counterparts, or
from an EM transient to identify candidate GW events. With its modular agentic architecture, 
GW-Eyes provides a flexible framework that is not limited to specific class of EM transients.

GW-Eyes can also assist human experts through natural-language interaction with
auxiliary tasks, enabling faster and more convenient information retrieval and
verification. Through natural-language queries alone,
GW-Eyes can report the basic properties of EM events, visualize the posterior
distribution of GW parameters, and rapidly determine whether a given coordinate
lies within a high-confidence region of a GW event.  In addition, as shown in
Figure~\ref{fig: total}B, the agentic design of GW-Eyes and its integration of
domain-specific tools provide clear advantages over general-purpose commercial
LLMs. For instance, GW-Eyes can build and maintain a local event database with
consecutive updates, while  supporting secure access to private data.  In the
highly specialized multi-messenger astronomy, general-purpose LLMs often
struggle to process domain-specific data effectively, whereas our integration of
dedicated tools substantially improves the adaptability of GW-Eyes.

\section*{EVALUATION AND APPLICATION}
\label{ sec:evaluation }

To evaluate the performance of GW-Eyes on counterpart association, as well as
the accuracy and efficiency of its tool use, we construct a simulated EM catalog
for Gravitational-Wave Transient Catalog 4.0
(GWTC-4.0)\textsuperscript{8} and test whether GW-Eyes can correctly
invoke relevant tools to recover the injected EM counterparts. 
Specifically, we select the first 50 GW events from GWTC-4.0. 
For each GW event, we inject one ground-truth counterpart satisfying the temporal, 
sky-position, and redshift criteria, together with two redshift-inconsistent 
confounders and ten temporally coincident unrelated events. 
All simulated EM events are then combined into a single catalog for retrieval tasks.
We evaluate GW-Eyes on two tasks: ``GW4EM'', which retrieves the injected EM 
counterpart for a given GW event, and ``EM4GW'', which retrieves the associated GW 
event for a given EM event. During evaluation, GW-Eyes can freely invoke tools to complete each task and is required to provide its final answer in a predefined format. 
Figure~\ref{fig: total}C shows that GW-Eyes achieves stable performance on association tasks 
across different LLM backends at relatively low cost.

We also demonstrate the application of GW-Eyes to EM counterpart searches using
real catalogs. Given that most events in GWTC-4.0 are BBH mergers, we take AGN
optical flares as an example to explore potential associations. We
use the Active Galactic Nuclei Flare Refined Catalog (AGNFRC),\textsuperscript{9}
whose temporal coverage overlaps with GWTC-4.0. We adopt a temporal window of
180 days after each GW trigger, apply GW-Eyes to search for potential temporal
and spatial associations between events in the two catalogs, and report
candidate counterparts whose sky-position falls within the 50\% and 90\%
credible regions of the GW localization. For events with reported redshifts, we
further require consistency within the 68\% credible region.  
The associations reported by GW-Eyes are independently cross-checked 
through manual execution of the tools, with consistent outcomes.
Results are shown in Figure~\ref{fig: total}D. Given the relatively long time window adopted and
poor sky localization of some GW events, temporal and spatial associations alone
can yield more than one EM candidate for certain GWs.  In addition, we compute
the number of associated GW-EM pairs ($N_{\rm GW-EM}$) with different GW
localization area thresholds, normalized by the number of EM events ($N_{\rm
EM}$) within the adopted 180-day post-trigger temporal windows of the 50
GWTC-4.0 events. Here, the GW localization area for each event is defined as the
sky area enclosed by the 90\% credible interval. Figure~\ref{fig: total}E shows
the cumulative ratio $N_{\rm GW-EM} / N_{\rm EM}$ as a function of the GW skymap
area threshold. In the future, the Einstein Telescope is expected to detect $\mathcal{O}(10^{4})$-$\mathcal{O}(10^{5})$ binary neutron-star mergers per year, and 
could enable the Rubin Observatory to detect ten to one hundred kilonovae annually.\textsuperscript{10} 
With the numbers of GW and EM events growing rapidly, counterpart association will become increasingly challenging. Fortunately, GW-Eyes can substantially help
address these challenges and support different groups in tailoring its use to
specific needs, greatly facilitating subsequent identification and follow-up
studies.


\section*{INTERPRETABILITY AND EXTENSIBILITY}
\label{ sec:discussion }

Compared to other machine-learning-based approaches, GW-Eyes, built on an LLM
agent framework, provides more interpretable reasoning traces and results.  On
the one hand, the full decision-making process, including tool usage and
parameters, is recorded in the human-readable natural language, facilitating
inspection and verification.  On the other hand, compared with general-purpose
LLMs, the integration of domain-specific tools significantly reduces
hallucinations in the specialized context of multi-messenger astronomy,
therefore improving the accuracy and reliability of its results.

GW-Eyes is highly extensible. In the present study, counterpart association
relies primarily on basic temporal and spatial overlap between GW and EM
signals, while more specialized criteria can be readily incorporated by
extending the agent's toolset. For different EM datasets, this may include
additional information such as spectra and images to support refined association
strategies. As data become increasingly structured, the agent's tool layer will continue to expand, while the advanced decision-making capabilities of LLMs will enable flexible workflows beyond simple if-else logic and provide substantial potential for further extension.


\section*{TOWARD AGENTIC MULTI-MESSENGER ASTRONOMY}
\label{ sec:conclu }

This {\it Letter} presents GW-Eyes, a first LLM-powered agent framework for GW
counterpart association in multi-messenger astronomy. Through the design of
Collector and Executor sub-agents and the integration of domain-specific tools,
GW-Eyes enables EM catalog management, counterpart association analyses, and
auxiliary support for human experts. Natural-language interaction significantly
improves flexibility and provides interpretable reasoning traces that facilitate
subsequent inspection and verification. In addition, GW-Eyes is highly
extensible, allowing more specialized association strategies to be readily
incorporated for different EM catalogs. We expect GW-Eyes to offer a promising
approach to alleviate the growing human workload in the big-data era of
multi-messenger astronomy.

\vspace{1em}

\textbf{REFERENCES}


\begin{enumerate}[label={[\arabic*]}]

    \item Abbott B. P., Abbott R., Abbott T. D. et al. (2017). Multi-messenger observations of a binary neutron star merger. \textit{Astrophys. J. Lett.} \textbf{848}:L12. DOI:10.3847/2041-8213/aa91c9

    \item Li L.-X. and Paczynski B. (1998). Transient events from neutron star mergers. \textit{Astrophys. J.} \textbf{507}:L59. DOI:10.1086/311680

    \item Abbott B. P., Abbott R., Abbott T. D. et al. (2017). A gravitational-wave standard siren measurement of the Hubble constant. \textit{Nature} \textbf{551}:85--88. DOI:10.1038/nature24471

    \item Bartos I., Kocsis B., Haiman Z. et al. (2017). Rapid and bright stellar-mass binary black hole mergers in active galactic nuclei. \textit{Astrophys. J.} \textbf{835}:165. DOI:10.3847/1538-4357/835/2/165

    \item Dax M., Green S. R., Gair J. et al. (2025). Real-time inference for binary neutron star mergers using machine learning. \textit{Nature} \textbf{639}:49--53. DOI:10.1038/s41586-025-08593-z

    \item Hosseinzadeh G., Paterson K., Rastinejad J. C. et al. (2024). SAGUARO: Time-domain infrastructure for the fourth gravitational-wave observing run and beyond. \textit{Astrophys. J.} \textbf{964}:35. DOI:10.3847/1538-4357/ad2170

    \item Wang H. and Zeng L. (2025). Automated algorithmic discovery for scientific computing through LLM-guided evolutionary search: A case study in gravitational-wave detection. \textit{arXiv} arXiv:2508.03661.

    \item Abac A. G., Abouelfettouh I., Acernese F. et al. (2025). GWTC-4.0: Updating the gravitational-wave transient catalog with observations from the first part of the fourth LIGO-Virgo-KAGRA observing run. \textit{Astrophys. J. Lett.} \textbf{1004}:L22. DOI:10.3847/2041-8213/ae2c74

    \item He L., Liu Z.-Y., Niu R. et al. (2025). A systematic search for active galactic nucleus flares in ZTF Data Release 23. \textit{Astrophys. J. Suppl. Ser.} \textbf{282}:13. DOI:10.3847/1538-4365/ae1d64

    \item Loffredo E., Hazra N., Dupletsa U. et al. (2025). Prospects for optical detections from binary neutron star mergers with the next-generation multi-messenger observatories. \textit{Astron. Astrophys.} \textbf{697}:A36. DOI:10.1051/0004-6361/202452863

\end{enumerate}

\vspace{1em}
\textbf{FUNDING AND ACKNOWLEDGMENTS}\\

This work is supported by the Beijing Natural Science Foundation (1242018), the
National Natural Science Foundation of China (12573042, 123B2043), the National
SKA Program of China (2020SKA0120300), the Max Planck Partner Group Program
funded by the Max Planck Society, and the High-performance Computing Platform of
Peking University. This research has made use of data or software obtained from the Gravitational
Wave Open Science Center (gwosc.org), a service of the LIGO Scientific
Collaboration, the Virgo Collaboration, and KAGRA. The funders had no role in study design, data collection and analysis, decision to publish, or preparation of the manuscript.

\vspace{1em}
\textbf{AUTHOR CONTRIBUTIONS}\\
Y.D. designed and developed the project code, performed the main data analyses, and drafted the manuscript. Y.K. contributed to discussions and analyses related to multi-messenger astronomy. 
J.Z. contributed to expertise in gravitational-wave counterpart associations.
X.Z. contributed to agent development. 
Z.W. contributed to expertise in gravitational-wave astronomy. 
L.S. contributed to the project design, the writing of the manuscript, and provided supervision and coordination of the whole project. 
All authors contributed to the manuscript and approved the final version.

\vspace{1em}
\textbf{DECLARATION OF INTERESTS}\\
The authors declare no conflicts of interest.\\

\vspace{1em}
\textbf{DATA AND CODE AVAILABILITY}\\
The code of GW-Eyes is publicly available via GitHub at
\url{https://github.com/Yiming-astro/GW-Eyes}. The repository provides
instructions for downloading the data involved in this {\it Letter}.


\clearpage

\begin{center}
    \textbf{\Large Supplemental Information}\\[2em]
\end{center}









\label{appendix:SI}

In this Supplementary Information, we provide 
a detailed description of the evaluation setup for GW-Eyes, 
extended discussions on the role of GW-Eyes and other LLM-based agents in multi-messenger astronomy, the supervision mechanism for the information retrieval process performed by Collector, and present three illustrative examples of GW-Eyes to better illustrate its reasoning and tool-calling processes.

\subsection*{Evaluation Setup}
\label{appendix_eval}

In this section, we provide a detailed description of the evaluation setup. 
The simulated EM catalog for evaluation is constructed as follows. 
We select the first 50 GW events from GWTC-4.0, spanning from May 18 to October 4, 2023. 
For each GW event, we inject one ground-truth EM counterpart occurring within 7 days after the GW trigger, 
located inside the 50\% credible region of the GW localization, and with a redshift in the 68\% interval 
of its distribution. We then add two confounding events that satisfy the temporal and sky-position criteria 
but have a false redshift, and ten unrelated events that satisfy only the temporal criterion. 
The simulated catalog contains 650 EM events and is used to test whether GW-Eyes correctly recovers the injection. 
We neglect localization uncertainties of EM events, since their sky positions are usually much more precise 
than those of GWs.

We evaluate GW-Eyes on two tasks: (i) ``GW4EM'', which retrieves the injected EM counterpart 
for a given GW event, and (ii) ``EM4GW'', which retrieves the associated GW event for a given EM event. 
GW4EM uses 50 GW events with injected counterparts, whereas EM4GW uses 50 EM events, including 20 events with 
corresponding GWTC-4.0 associations, 15 simulated confounding events, and 15 unrelated events. For the latter 
two groups, no associated GW event is expected in GWTC-4.0. During evaluation, GW-Eyes is allowed to freely 
invoke tools to complete each task and is required to provide its final answer in a predefined format. 
Performance stability is assessed by comparing the final answer with the ground truth. 
For each query, the answer provided by GW-Eyes is compared with the injected ground-truth association, which may correspond either 
to a specific event or to no associated event. 
Across different LLM backends, GW-Eyes correctly completes all association tasks, 
demonstrating stable performance at relatively low cost.

\subsection*{Extended Discussions}
\label{appendix_discussions}

In this section, we further discuss the potential roles of GW-Eyes and other LLM-based agents in future multi-messenger astronomy. 
By drawing an analogy with traditional multi-messenger astronomy workflows, we clarify the respective roles of LLMs and tools within an agentic framework. 
We also discuss how agentic frameworks may deliver greater value in a future era of increasingly standardized data.

\begin{figure}[t]
    \centering
    \includegraphics[width=0.88\textwidth]{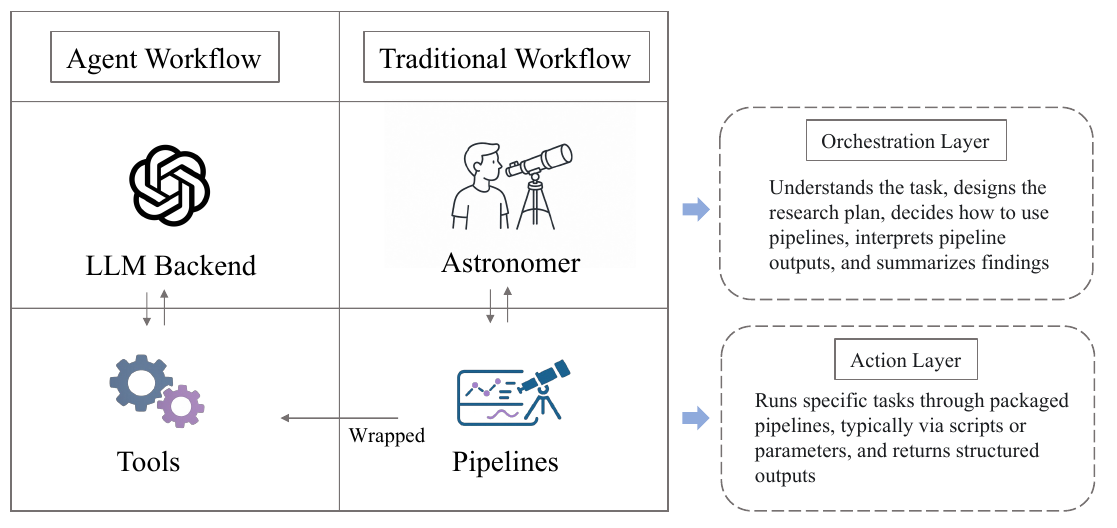}
    \caption{\textbf{Comparison between the agent workflow and the traditional workflow.} The LLM backend and the astronomer both serve as the orchestration layer, completing the task through interactions with action layer; standardized pipelines constitute the action layer and can be wrapped as agent tools.}
\label{fig: analogy}
\end{figure}

As illustrated in Figure~\ref{fig: analogy}, by comparison with the conventional workflow in which human astronomers conduct research using standardized data-analysis pipelines, we further clarify the roles played by the different components of an agentic system such as GW-Eyes.
Both the LLM backend and the human astronomer serve as an ``orchestration layer'', responsible for understanding the task, formulating a research plan, determining how pipelines should be invoked, and summarizing the findings. 
The agent toolkit and conventional pipelines constitute the ``execution layer'': they receive scripts or parameters, carry out specific subtasks, and return structured outputs. 
Importantly, standardized pipelines naturally serve as executable tools and can be readily wrapped as tools for an LLM agent. 
In this sense, agentic frameworks are not intended to compete with standardized pipelines. 
Rather, they are designed to reduce the effort human astronomers spend on repetitive tool invocation, allowing them to focus on scientific discoveries.

With the anticipated surge in the data volume of multi-messenger astronomy, LLM agents are expected to play an increasingly important role.
For tasks that involve simple logical decisions or require extremely low latency, standardized pipelines provide timely responses, whereas for tasks that involve complex reasoning, decision-making, or long-horizon analysis, LLM agents provide more flexible information integration and analytical decision support.
Compared with standardized pipelines, an LLM agent can make use not only of structured data in standard formats, but also important information expressed in natural language.
This includes scientific communications in notices that are difficult to encode in structured form, such as observing conditions and interpretations from the observing teams.
LLM agents may therefore offer stronger capabilities for extracting and integrating heterogeneous information.
In addition, LLM agents provide with a natural-language interface for data analysts. Compared with conventional workflows that rely heavily on code and scripts, this mode of interaction may further improve the efficiency of astronomers and lower the barrier to enter multi-messenger astronomy, thereby facilitating the development of the field.

Moreover, GW-Eyes offers strong flexibility for future extensions.
Benefiting from the general capabilities demonstrated by LLMs, new functions and tools can often be incorporated without training the LLMs themselves. 
Instead, such extensions can be achieved by improving the design of the agentic loop, providing reference examples, and developing agent skills. 
This extensibility could further broaden the impact of the agentic framework in future studies.

\subsection*{Supervision mechanism for the information retrieval}
\label{appendix_supervision}

\begin{figure}[t]
    \centering
    \includegraphics[width=0.6\textwidth]{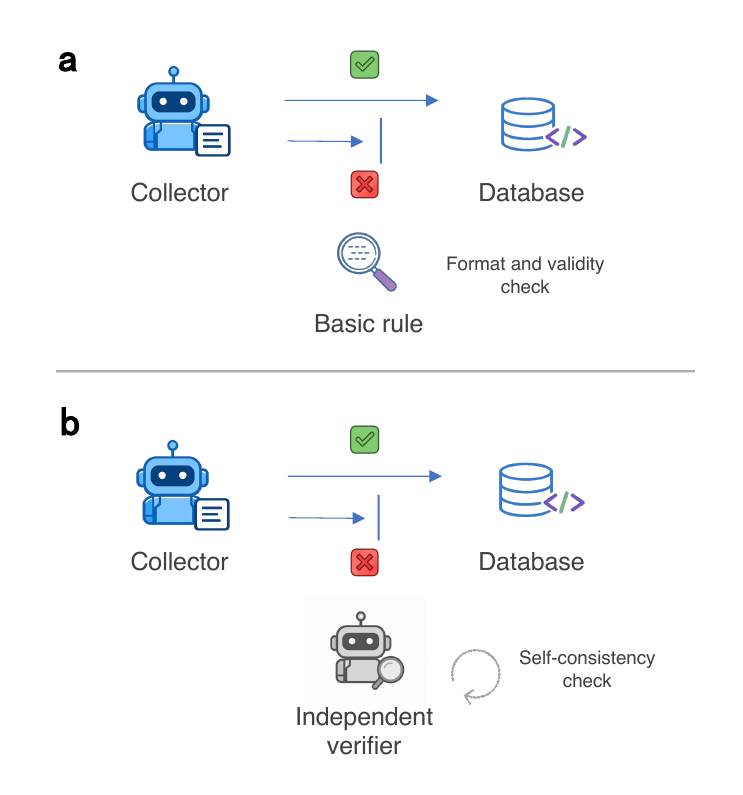}
    \caption{\textbf{Schematic illustration of the supervision mechanisms for information retrieval in the Collector.} (a) Basic rule-based checks for validating the extracted information. (b) An LLM-based independent verifier that uses consistency across repeated extractions.}
\label{fig: supervision}
\end{figure}

To ensure the accuracy of the data collected by the Collector, we designed a set of supervision mechanisms for its information retrieval process.
The first supervision mechanism consists of \textit{basic rule-based checks}, as illustrated in Figure~\ref{fig: supervision}a, which verify the formatting and basic plausibility of the extracted data. 
For example, if a negative redshift value is extracted, the Collector rejects the corresponding entry, prevents it from being written to the database, and raises an alert. These rule-based checks prevent clearly invalid data from entering the database. The second supervision mechanism introduces an additional LLM-based independent verifier, which extracts key information such as right ascension, declination, and redshift independently by the agent (see Figure~\ref{fig: supervision}b). 
This verification process can be repeated multiple times to produce several independent extraction results. 
If any inconsistencies or potential sources of misinterpretation are found in the results, the Collector rejects the entry, prevents it from being written to the database, and raises an alert. 
The scope and rigor of these monitoring and verification mechanisms can be further strengthened through system-level instructions.
Only when repeated independent verifications yield consistent results does the Collector write the information to the database. 
This mechanism further reduces the uncertainty associated with a single model extraction and provides stronger safeguards for data quality.

\subsection*{Illustrative Examples of GW-Eyes}
\label{appendix_examples}

\begin{figure}[t]
    \centering
    \includegraphics[width=0.98\textwidth]{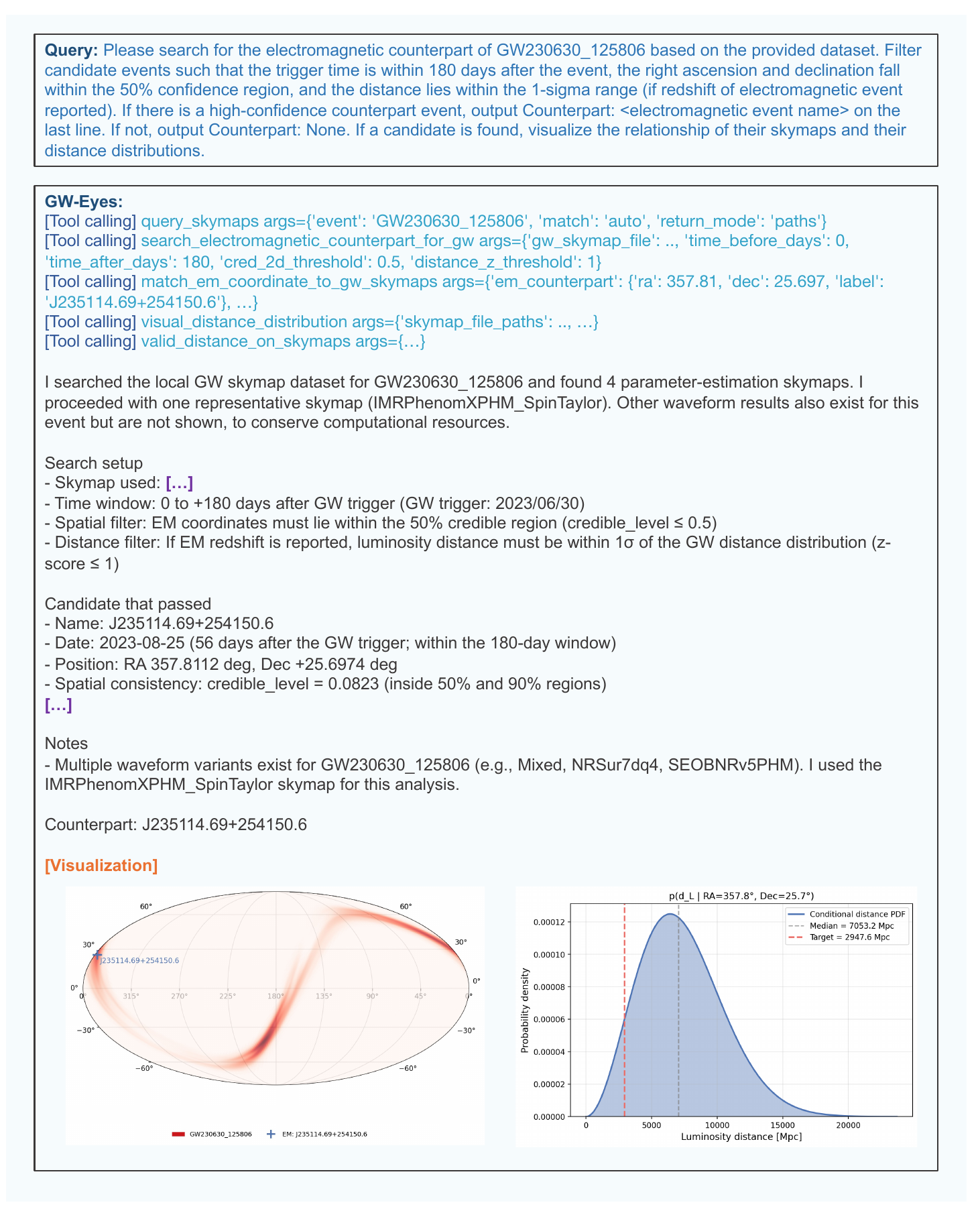}
    \caption{\textbf{An example of GW-Eyes applied to the counterpart
    association task for BBH mergers.} Upon receiving a counterpart association
    task, GW-Eyes retrieves the relevant GW information from the database,
    interprets the requested criteria, invokes domain-specific tools for EM
    candidate matches, and finally produces a detailed report along with the
    corresponding visualizations. Note that the EM catalog used here is AGNFRC.}
\label{fig: example1}
\end{figure}

\begin{figure}[t]
    \centering
    \includegraphics[width=0.98\textwidth]{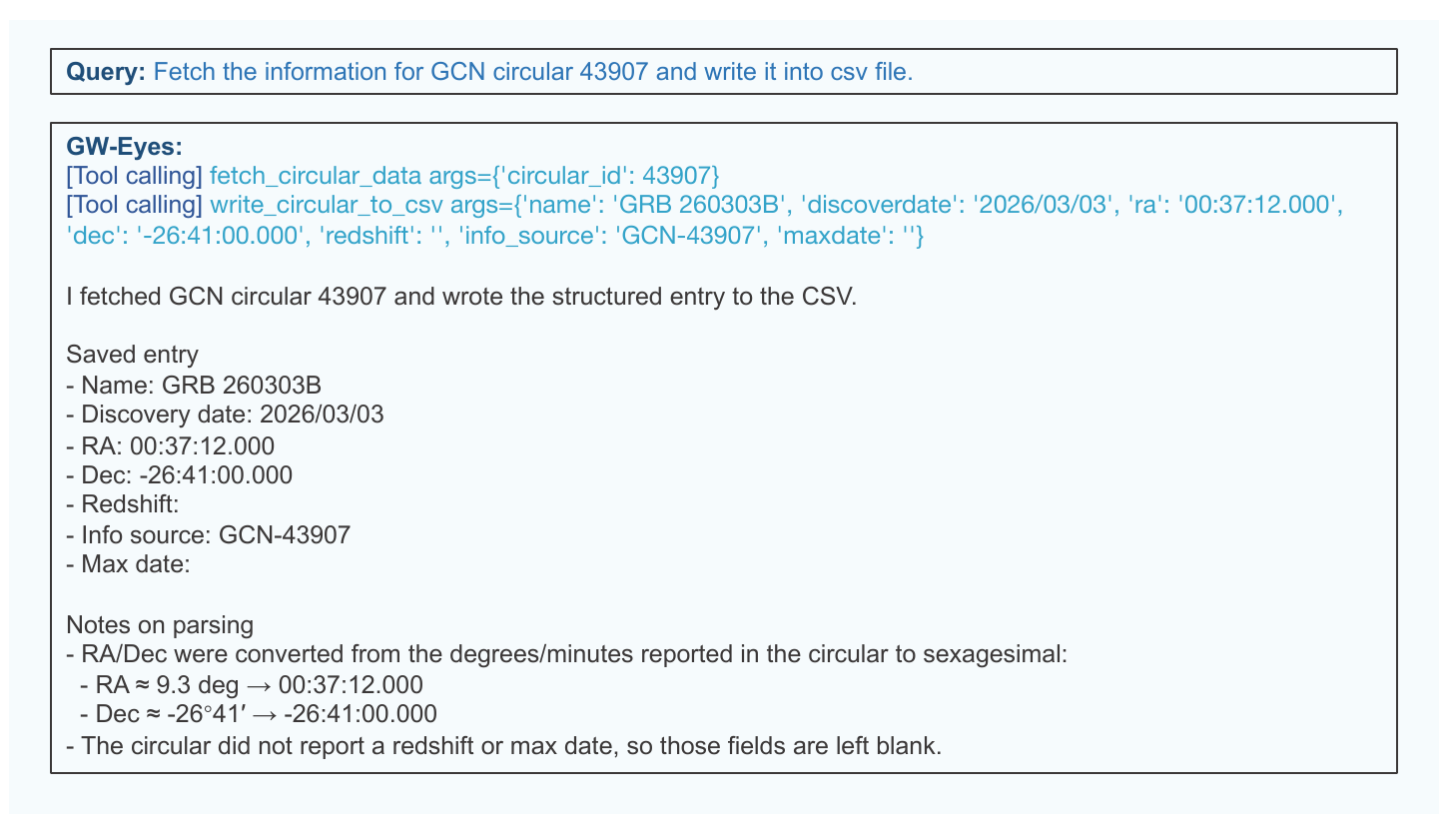}
    \caption{\textbf{An example of GW-Eyes for updating catalogs from GCN
    circulars.} Upon receiving a request to update catalogs, GW-Eyes retrieves
    the relevant information from the web, extracts key entries, invokes tools
    according to a predefined workflow to update the database, and presents the
    corresponding results.}
\label{fig: example2}
\end{figure}

\begin{figure}[t]
    \centering
    \includegraphics[width=0.98\textwidth]{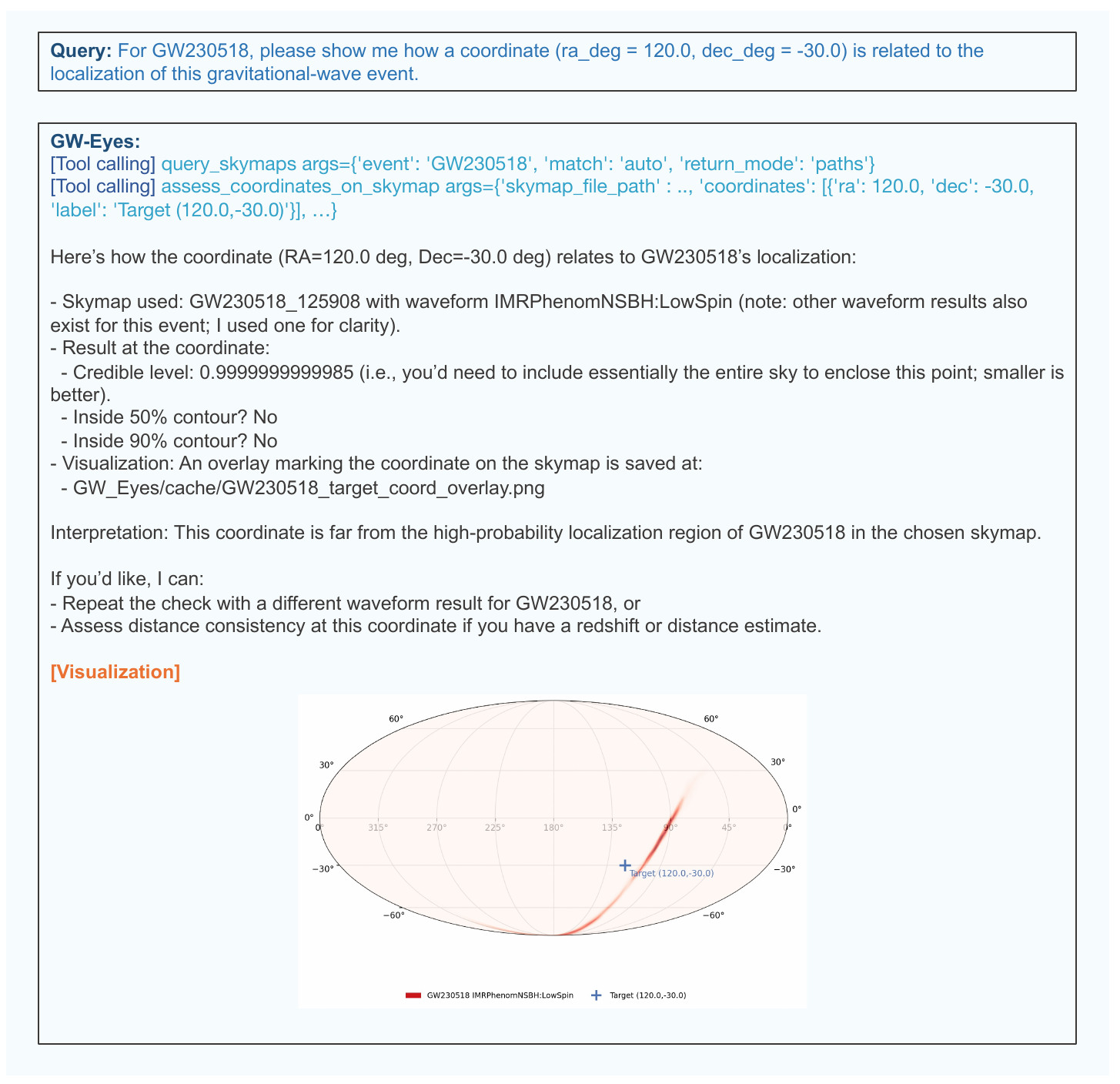}
    \caption{\textbf{An example of GW-Eyes for rapid verification tasks.} For
    quickly assessing the relationship between a given sky coordinate and the
    localization of a GW event, GW-Eyes can invoke specialized tools and return
    intuitive visualizations as well as detailed analysis results. }
\label{fig: example3}
\end{figure}

In this section, we present three examples of using GW-Eyes,
including counterpart-association tasks for BBH mergers (Figure~\ref{fig:
example1}), catalog updates based on GCN Circulars (Figure~\ref{fig: example2}),
and rapid verification tasks (Figure~\ref{fig: example3}).  These examples
demonstrate the tool-calling procedures and outputs of GW-Eyes.  For brevity,
details such as file paths and portions of the reasoning process are omitted.

\end{document}